\documentclass[12pt, draftclsnofoot, onecolumn]{IEEEtran}
\ifCLASSINFOpdf
\else
\fi
\usepackage{graphicx}
\usepackage{epstopdf}
\usepackage{setspace}
\usepackage{amsmath}
\usepackage{amssymb} 
\usepackage{bm} 
\usepackage{cite}
\usepackage{float}
\usepackage[usenames,dvipsnames]{pstricks}
\usepackage{epsfig}
\usepackage{pst-grad} 
\usepackage{pst-plot} 
\usepackage{tabularx}
\usepackage[keeplastbox]{flushend}
\usepackage[T1]{fontenc}
\newcommand{\lb} {\left}
\newcommand{\rb} {\right}
\newcommand{\nn} {\nonumber}

\allowdisplaybreaks
\flushend
\usepackage{caption}
\usepackage{subcaption}
\usepackage[utf8]{inputenc}

\IEEEaftertitletext{\vspace{-2.4\baselineskip}}

\begin{document}
\title{Ergodic Secrecy Rate of Optimal Source Selection in a Multi-Source System with Unreliable Backhaul}
 \author{Chinmoy Kundu,~\IEEEmembership{Member,~IEEE,} and Mark F. Flanagan,~\IEEEmembership{Senior Member, IEEE}
\thanks{Chinmoy Kundu and Mark F. Flanagan are with University College Dublin, Ireland (email: chinmoy.kundu@ucd.ie and mark.flanagan@ieee.org).}
\thanks{This publication has emanated from research conducted with the financial support of Science Foundation Ireland (SFI) under Grant Number 17/US/3445.
} 
}

\maketitle
 \thispagestyle{empty}
 \pagestyle{empty}
\pagestyle{plain} 
 \begin{abstract}
The use of multiple source nodes with wireless backhaul is considered for secrecy enhancement through source node selection in future wireless networks. The ergodic secrecy rate (ESR) of  {optimal source node selection in the presence of multiple eavesdroppers} over independent non-identically distributed (INID) Rayleigh fading channels is evaluated in closed-form. At high signal-to-noise ratio (SNR),  {the ESR is expressed as a simple weighted summation where each term relates to the contribution of an individual source and eavesdropper}. An asymptotic analysis shows the effect of the system parameters and backhaul reliability on the performance.  {The proposed method can provide a generalized solution for the ESR of optimal \textit{transmit antenna} selection in multi-antenna systems, optimal \textit{source  node} selection, and optimal \textit{relay selection} with or without unreliable backhaul.} 
\end{abstract}
\vspace{-.4cm}
\section{Introduction}
\label{sec_intro}
 {Transmit node selection (TS) is an  attractive way to achieve diversity gain in distributed, dense, heterogeneous, and low-complexity future wireless networks where deploying multiple-antenna transmitters may be impractical.}
TS for physical layer security has been extensively studied in 
\cite{Kundu_relsel_TWC15, Kundu_selection_GC16,  Shao_Modify_and_Forward, Kundu_TVT19, kim_Poor_Secrecy_full-duplex, kundu_LSTM_GC20,  kim2015security,  kim2016secrecy, kim_Poor_Secrecy_Finite-Sized, Kim_Poor_Secrecy_CDD } and references therein. 
 Further, as a cost-effective solution against deploying large-scale wired backhaul in future networks, wireless backhaul has been proposed 
\cite{Kundu_TVT19,kundu_LSTM_GC20,kim2015security,kim2016secrecy,kim_Poor_Secrecy_Finite-Sized,kim_Poor_Secrecy_full-duplex,Kim_Poor_Secrecy_CDD}. These papers study the effect of wireless backhaul reliability on the secrecy performance of TS schemes. Various cooperative relay selection (RS) schemes were also developed for secrecy without backhaul in \cite{Kundu_relsel_TWC15,Kundu_selection_GC16,Shao_Modify_and_Forward}.  Ergodic secrecy rate (ESR) was analyzed for various network configurations over flat-fading \cite{Kundu_TVT19, Shao_Modify_and_Forward,  kim_Poor_Secrecy_full-duplex} and frequency selective fading channels 
\cite{kim2016secrecy,kim_Poor_Secrecy_Finite-Sized,Kim_Poor_Secrecy_CDD,kim2015security}.  

TS and RS problems with or without wireless backhaul are analytically similar to their predecessor, transmitter antenna selection (TAS) \cite{Sadeque_Subramanian_Average_secrecy_rate}. As backhaul reliability is defined by the link failure probability, the assumption of perfect backhaul leads to TAS or RS problems \cite{kim2015security}. RS problems are similar to TS problems with unreliable backhaul when the relay forwarding strategy is conditional on the first-hop link, as in \cite{Kundu_selection_GC16}. RS problems are also similar to TAS problems when end-to-end performance is limited by the worst link with decode-and-forward (DF) relaying, as in \cite{Kundu_relsel_TWC15}. 

 {
The optimal selection strategy maximizes the secrecy rate of the system considering global channel knowledge. Optimal TS was analyzed in \cite{Kundu_relsel_TWC15,Kundu_selection_GC16, kundu_LSTM_GC20}; however, ESR was not evaluated. 
ESR was obtained for sub-optimal TS including backhaul uncertainty in \cite{kim_Poor_Secrecy_full-duplex, kim2015security, Shao_Modify_and_Forward,kim2016secrecy,kim_Poor_Secrecy_Finite-Sized}; however, this can not provide the maximum secrecy rate. 
ESR is only available in a computable form for the simplest case of independent and identically distributed (IID) fading channels in \cite{Kundu_TVT19,Sadeque_Subramanian_Average_secrecy_rate}. Also, although \cite{Kundu_TVT19} included backhaul uncertainty into the analysis, this was not done in   \cite{Sadeque_Subramanian_Average_secrecy_rate}. 
}

 {Despite the similarity between TAS, TS, and RS with or without unreliable backhaul, the ESR of the optimal selection strategy is not available in any of these cases in closed-form.
Motivated by this, we provide a closed-form solution for the ESR of the optimal TS strategy with wireless backhaul in the presence of multiple eavesdroppers. } We also analyze a scheme for TS on the basis of the ratio of instantaneous signal-to-noise ratios (SNRs) of the destination and eavesdropper channels, which leads to a high-SNR approximation of the ESR for optimal selection. Based on this approximation, we provide a closed-form solution for the ESR as well as its asymptotic expression.  { This provides a generalized solution for the ESR of optimal TAS, optimal TS, and optimal RS with DF relaying, with or without wireless backhaul.}
 

\textit{Notation:} $\mathbb{P}[\cdot]$ denotes probability, $\mathbb{E}[\cdot]$ denotes expectation, $F_{X} (\cdot)$ denotes the cumulative distribution function (CDF) of a random variable (RV) $X$, and $f_{X} (\cdot)$ represents the corresponding probability density function (PDF). $[K]$ denotes the set $\{ 1, 2, \ldots, K \}$, and $|\mathcal{A}|$ denotes the cardinality of the set $\mathcal{A}$. Finally, vectors are denoted by bold type. 
\vspace{-.3cm}
\section{System Model}
\label{sec_system}
 {The system consists of an access point $A$, $K$ source nodes $S_k$ for $k \in [K]$, a destination node $D$, and $N$ eavesdroppers $E_n$ for $n \in [N]$. Each of these nodes is equipped with a single antenna. $A$ provides a wireless backhaul connection to each $S_k$.
For any $k$, the channels $S_k$-$E_n$ (for $n \in [N]$) are assumed to be IID, while all other channels are assumed to be independent and non-identically distributed (INID).
All individual links are assumed to exhibit frequency-flat Rayleigh fading, and thus the corresponding channel SNRs are exponentially distributed.
The exponential channel parameter for the $S_k$-$D$ link is denoted by $\beta_k$, while that for the $S_k$-$E_n$ link (for any $n$) is denoted by $\alpha_k$. Therefore, $1/\beta_k$ and $1/\alpha_{k}$ are the average received SNRs, respectively.
We consider the scenario of non-colluding eavesdroppers. In this case, if source $S_k$ is transmitting, the secrecy performance is determined by the eavesdropper $E_n$ with maximum (instantaneous) SNR of the $S_k$-$E_n$ link. This maximum SNR, denoted by $\gamma_{E_k}$, has PDF given by
\vspace{-.3cm}
\begin{align}
\label{eq_eavesdropper_sc}
 &f_{\gamma_{E_k}}(x)=
\sum_{n=1}^{N}\psi_nn\alpha_{k}\exp(-n\alpha_{k}x),
 \end{align} 
where $\psi_n=(-1)^{n+1}\binom{N}{n}$.}
The uncertainty of the $k$th wireless backhaul link is modelled by a Bernoulli random variable {\cite{Kundu_TVT19}}; the probability that the $k$th backhaul link is active is equal to $\delta_k$ and the probability that it is inactive is equal to $1-\delta_k$, where $0 \le \delta_k \le 1$. 
The optimal source (for secrecy improvement) is selected assuming global knowledge of CSI and backhaul activity, i.e., the optimal source is selected from the set of sources whose backhaul links are active. The assumption of availability of global CSI is common in the secrecy literature (c.f. \cite{kim2015security,Kundu_TVT19, Kundu_relsel_TWC15,Kundu_selection_GC16,kundu_LSTM_GC20} and references therein); moreover, this assumption can provide the best achievable performance bound. 

 {Considering that $\mathcal{S} \ne \emptyset$ is the subset of source nodes with active backhaul links, the corresponding optimal secrecy rate $C_s^{\mathcal{S}}$ in nats per channel use (npcu) is given by}
  {\begin{align}
  \label{eq_set_capa}
  C_s^{\mathcal{S}} = \max\{\ln(\Gamma_{\mathcal{S}}),0\},
   \end{align}
    where \vspace{-.2cm}
  \begin{align}
  \label{eq_max_snr_ratio}
  \Gamma_{\mathcal{S}}=\max_{ k\in\mathcal{S}}\Big\{\frac{1+\gamma_{D_k}}{1+\gamma_{E_{k}}}\Big\},
    \end{align}
    and $\gamma_{D_k}$ is the received SNR for the channel from $S_k$ to $D$. 
Then the achievable secrecy rate, $C_s$, of the system can be evaluated by taking into account all possible active backhaul sets $\mathcal{S}$, i.e.,
  \vspace{-.2cm}
  \begin{align}
\label{eq_capacity}
C_s=\sum\limits_{\mathcal{S} \subseteq [K]} \mathbb{P} [\mathcal{S}]  C_s^{\mathcal{S}} ,
\end{align}
where $\mathbb{P}[\mathcal{S}]$ denotes the probability of occurrence of subset $\mathcal{S}$.
Here we assume that when $\mathcal{S}$ is the empty set (no active backhaul links), the corresponding $C_s^{\mathcal{S}}$ is equal to $0$. 
The probability of the source node set $\mathcal{S}$ having active backhauls can be expressed as}
 {\begin{align}
\label{eq_prob_sel}
  \mathbb{P} [\mathcal{S}] = \prod_{k\in \mathcal{S} } \delta_k \prod_{j \notin {\mathcal{S}}} (1-\delta_j),
  \end{align}
and the ESR of the system is then evaluated using \eqref{eq_capacity} as
\begin{align}
\label{eq_avg_capacity}
\bar{C}_s=\sum\limits_{\mathcal{S} \subseteq [K]}    \mathbb{P} [\mathcal{S}] \mathbb{E}[C_s^{\mathcal{S}}]=\sum\limits_{\mathcal{S} \subseteq [K]}    \mathbb{P} [\mathcal{S}]  \int_1^\infty \ln(x)f_{\Gamma_{\mathcal{S}}}(x)dx.
\end{align}
}

 \vspace{-.7cm}
\section {ESR of Optimum Source Selection}
\label{sec_avg_sec}
 {
In this section, we will find the ESR of the optimal source selection scheme defined in  (\ref{eq_avg_capacity}). Our approach will be to find the distribution of $\Gamma_{\mathcal{S}}$ in (\ref{eq_max_snr_ratio}) in order to evaluate (\ref{eq_avg_capacity}). }
 \vspace{-.3cm}
 {\subsection{Distribution of $ \Gamma_{\mathcal{S}}$}
It can be seen from (\ref{eq_set_capa}) that $C_s^{\mathcal{S}} = 0$ whenever  $\Gamma_{\mathcal{S}} \le 1$. Therefore, it will be sufficient to characterize the CDF $F_{\Gamma_{\mathcal{S}}}(x)$ only for the case where $x > 1$. This CDF may be evaluated 
with the help of the exponential CDF of $\gamma_{D_k}$ and the PDF of $\gamma_{E_k}$ from (\ref{eq_eavesdropper_sc}) as  
\begin{align}
  &F_{\Gamma_{\mathcal{S}}}(x)=\mathbb P[\Gamma_{\mathcal{S}}\le x]
  =\prod_{k\in \mathcal{S}} \mathbb P\Big[\frac{1+\gamma_{D_k}}{1+\gamma_{E_k}}\le x\Big]\nn\\
\label{eq_product}
&=\prod_{k\in \mathcal{S}}\Big[1-\sum_{n=1}^{N}\psi_n\frac{na_k c_k \exp(-\beta_k x)}{x+na_k}\Big], ~~\text{for}~~ x>1,
\end{align}
where $a_k={\alpha_k}/{\beta_k}$ and $c_k=\exp(\beta_k)$.
We express the product in \eqref{eq_product} as a series via
\begin{align}
\label{eq_cdf_os}
&F_{\Gamma_{\mathcal{S}}}(x)=\sum\limits_{\mathcal{M} \subseteq \mathcal{S}} \lb(-1\rb)^{|\mathcal{M}|}e^{-\beta_{\mathcal{M}}x}
\prod_{k\in\mathcal{M}}\Big[\sum_{n=1}^{N}\frac{\psi_nna_{k}c_{k}}{x+na_{k}}\Big]
\nn\\ 
&=\sum\limits_{\mathcal{M} \subseteq \mathcal{S}} \lb(-1\rb)^{|\mathcal{M}|}e^{-\beta_{\mathcal{M}}x}\sum_{\substack{\mathbf {i} \in [N]^{|\mathcal{M}|}\\\mathbf{i}=(i_k)_{k\in\mathcal{M}}}}\prod_{k \in \mathcal{M}}\frac{\psi_{i_k} i_ka_{k}c_{k}}{x+i_ka_{k}},
\end{align}
where  $\beta_{\mathcal{M}}=\sum_{k\in\mathcal{M}}\beta_{k}$. 
The corresponding PDF, $f_{\Gamma_{\mathcal{S}}}(x)$, can be evaluated by differentiating \eqref{eq_cdf_os}. However, when deriving the expected value of \eqref{eq_set_capa} we will use integration by parts, which is simpler and does not require derivation of the PDF.}
\vspace{-.3cm}
 {
\subsection{Evaluation of ESR}
The ESR for the active backhaul set $\mathcal{S}$ may be obtained by taking the expected value of (\ref{eq_set_capa}) using the integral shown in (\ref{eq_avg_capacity}) via (\ref{eq_cdf_os}), and using integration by parts to obtain 
\begin{align}
\label{eq_averaging}
&\mathbb E[C_s^{\mathcal{S}}]=
\int_1^\infty \ln(x)f_{\Gamma_{\mathcal{S}}} (x)dx\nn\\
&=\lb[\ln(x)F_{\Gamma_{\mathcal{S}}}(x)\rb]_1^\infty-
\sum\limits_{\mathcal{M} \subseteq \mathcal{S}} \lb(-1\rb)^{|\mathcal{M}|}
\nn\\
&\sum_{\substack{\mathbf {i} \in [N]^{|\mathcal{M}|}\\\mathbf{i}=(i_k)_{k\in\mathcal{M}}}} \int_1^\infty\frac{\exp{(-\beta_{\mathcal{M}}x)}\prod_{k \in \mathcal{M}}\psi_{i_k} i_ka_{k}c_{k}}{x\prod_{k \in \mathcal{M}}\lb(x+i_ka_{k}\rb)}
 dx.
\end{align}
By taking the limit, it can be easily shown that the first term is equal to zero. Furthermore, we assume that the values $i_ka_k$ are all different\footnote{Note that this assumption holds with overwhelming probability. In the special case where some or all $i_ka_k$ are the same, the partial fractions need to be evaluated differently. Since these cases are rather tedious to enumerate, these calculations are omitted.} to evaluate (9) with the help of partial fractions. More formally, we assume that the channel parameters are such that for any $p,q \in [N]$ and any  $j,k \in [K]$ with $j \ne k$, we have $p a_k \ne q a_j$.
Using the method of partial fractions, the integration in the second term can be solved as
\begin{align}
\label{eq_averaging2}
&\mathbb E[C_s^{\mathcal{S}}]
=-\sum\limits_{\mathcal{M} \subseteq \mathcal{S}} \lb(-1\rb)^{|\mathcal{M}|}
\sum_{\substack{\mathbf {i} \in [N]^{|\mathcal{M}|}\\\mathbf{i}=(i_k)_{k\in\mathcal{M}}}}\Big(\prod_{k \in \mathcal{M}}\psi_{i_k} i_ka_{k}c_{k}\Big)\times \nn\\
 &\int_1^\infty \Big( \frac{A_{\mathbf{i}}^{(0)} }{x}+ \sum_{k\in\mathcal{M}} \frac{A_{\mathbf{i}}^{(k)} }{x+i_ka_k}\Big)\exp{(-\beta_{\mathcal{M}}x)}dx\nn\\
 &=\sum\limits_{\mathcal{M} \subseteq \mathcal{S}} \lb(-1\rb)^{|\mathcal{M}|}
\sum_{\substack{\mathbf {i} \in [N]^{|\mathcal{M}|}\\\mathbf{i}=(i_k)_{k\in\mathcal{M}}}}\Big(\prod_{k \in \mathcal{M}}\psi_{i_k} i_ka_{k}c_{k}\Big)\Big(A_{\mathbf{i}}^{(0)}\text{Ei}(- \beta_{\mathcal{M}})\nn\\
&+\sum_{k\in\mathcal{M}} A_{\mathbf{i}}^{(k)} \exp{(\beta_{\mathcal{M}} i_ka_{k})}\text{Ei}\lb(-\beta_{\mathcal{M}}(1+i_ka_{k})\rb)\Big),
\end{align}
where $A_{\mathbf{i}}^{(0)}$ and $A_{\mathbf{i}}^{(k)}$ for $k\in\mathcal{M}$ are found by the method of partial fractions as
\begin{align}
A_{\mathbf{i}}^{(0)}=\frac{1}{\prod\limits_{k\in\mathcal{M}} i_ka_{k}},
A_{\mathbf{i}}^{(k)} =  \frac{(-1)^{|\mathcal{M}|}}{i_ka_{k}\prod\limits_{\substack{j\in\mathcal M\\j\ne k}} (i_ka_{k}-i_ja_{j  })}.\nn  
\end{align}
The closed-form solution \eqref{eq_averaging2} is obtained using the integral solutions 3.351.5 and 3.352.2 in \cite{book_ryzhik}, respectively.
}

 {
By substituting \eqref{eq_averaging2} into \eqref{eq_avg_capacity} and using \eqref{eq_prob_sel}, the ESR for optimal source selection with backhaul uncertainty can be derived in closed-form.} 










\vspace{-.2cm}
\section{High-SNR Analysis}
\label{sec_opt_high}
In this section we derive an approximate expression for the ESR which is accurate when both the main channel and eavesdropper channel operate in the high-SNR regime, i.e., $\gamma_{D_k}, \gamma_{E_k} \gg 1$ for all $k$, to obtain an approximate expression which is simpler than \eqref{eq_averaging2}. At high SNR, ${\Gamma_{\mathcal{S}}}$ can be approximated by neglecting unity from both the numerator and the denominator in (\ref{eq_max_snr_ratio}) to obtain
${\Gamma_{\mathcal{S}}} \approx \max_{k\in \mathcal{S}}\lb\{
\frac{\gamma_{D_k}}{\gamma_{E_k}}\rb\}$.
This effectively states that at high SNR, the optimal selection is on the basis of maximizing the SNR ratio between the destination and eavesdropper channels. We will first proceed to find the CDF of $\Gamma_{\mathcal{S}}$.

\vspace{-.4cm}
 {
\subsection{Distribution of ${\Gamma_{\mathcal{S}}}$ with high-SNR approximation}
For $\mathcal{S}\ne\emptyset$, the CDF of ${\Gamma_{\mathcal{S}}}$ at high SNR is evaluated following a similar approach to (\ref{eq_product}) as 
\begin{align}
\label{eq_cdf_inid_high1}
&F_{\Gamma_{\mathcal{S}}}(x)  
=\prod_{k\in\mathcal{S}} \Big(1-\sum_{n=1}^{N}\frac{\psi_nna_k}{x+na_k}\Big)
\\
&=\sum\limits_{\mathcal{M} \subseteq \mathcal{S}} \lb(-1\rb)^{|\mathcal{M}|}\prod_{k\in\mathcal{M}}\Big( \sum_{n=1}^{N}\frac{\psi_n na_k}{x+na_k}\Big)\nn\\
&=\sum\limits_{\mathcal{M} \subseteq \mathcal{S}} \lb(-1\rb)^{|\mathcal{M}|}\sum_{\substack{\mathbf {i} \in [N]^{|\mathcal{M}|}\\\mathbf{i}=(i_k)_{k\in\mathcal{M}}}}\Big(\prod\limits_{k \in \mathcal{M}} \frac{\psi_{i_k}i_ka_{k}}{x+i_ka_{k}}\Big)\nn\\
\label{eq_cdf_inid_high2}
&=\sum\limits_{\mathcal{M} \subseteq \mathcal{S}} \lb(-1\rb)^{|\mathcal{M}|}\sum_{\substack{\mathbf {i} \in [N]^{|\mathcal{M}|}\\\mathbf{i}=(i_k)_{k\in\mathcal{M}}}} \sum_{k\in\mathcal{M}}\frac{A_{\mathbf{i}}^{(k)}\prod\limits_{j\in\mathcal{M}}\psi_{i_j}i_ja_{j}}{\lb(x+i_ka_{k}\rb)},
\end{align}
where $A_{\mathbf{i}}^{(k)}$ is evaluated by the method of partial fractions as
\begin{align}
\label{eq_partial_frac_high}
A_{\mathbf{i}}^{(k)}=  \frac{(-1)^{|\mathcal{M}|+1}}{\prod_{\substack{j\in\mathcal M\\ j \ne k}} (i_ka_{k}-i_ja_{j})}.
\end{align}
In contrast to the analysis of Section~\ref{sec_avg_sec}, in the high-SNR scenario it is easier to find the ESR by direct integration using the PDF of $\Gamma_{\mathcal{S}}$. This PDF may be evaluated using (\ref{eq_cdf_inid_high2}) as
 \begin{align}
\label{eq_pdf_high_snr}
f_{\Gamma_{\mathcal{S}}}(x)&=\sum\limits_{\mathcal{M} \subseteq \mathcal{S}} \lb(-1\rb)^{|\mathcal{M}|+1}\sum_{\substack{\mathbf {i} \in [N]^{|\mathcal{M}|}\\\mathbf{i}=(i_k)_{k\in\mathcal{M}}}}\sum_{k\in\mathcal{M}}\nn\\
& \Big(\prod\limits_{k\in\mathcal{M}}\psi_{i_k}i_ka_{k}\Big)\frac{A_{\mathbf{i}}^{(k)}}{\lb(x+i_ka_{k}\rb)^2}.
\end{align}
}

 {
\subsection{Evaluation of ESR with high-SNR approximation}
The ESR corresponding to the active backhaul set $\mathcal{S}$ in the high-SNR scenario is evaluated with the aid of \eqref{eq_pdf_high_snr}; proceeding similarly to (\ref{eq_averaging}), we obtain
\begin{align}
\label{eq_average_highsnr1}
&\mathbb E[C_s^{\mathcal{S}}]
=\sum\limits_{\mathcal{M} \subseteq \mathcal{S}} \lb(-1\rb)^{|\mathcal{M}|+1}\sum_{\substack{\mathbf {i} \in [N]^{|\mathcal{M}|}\\\mathbf{i}=(i_k)_{k\in\mathcal{M}}}} \sum_{k\in\mathcal{M}}\Big(\prod_{k\in\mathcal{M}}\psi_{i_k}i_ka_{k}\Big)\nn\\
&\int_1^\infty\frac{A_{\mathbf{i}}^{(k)}\ln(x) }{\lb(x+i_ka_{k}\rb)^2}dx
=\sum\limits_{\mathcal{M} \subseteq \mathcal{S}} \sum_{\substack{\mathbf {i} \in [N]^{|\mathcal{M}|}\\\mathbf{i}=(i_k)_{k\in\mathcal{M}}}}
 \sum_{k\in\mathcal{M}}\nn\\
&\Big(\prod\limits_{\substack{j\in\mathcal M\\ j \ne k}} \frac{\psi_{i_j}i_ja_{j}}{(i_ka_{k}-i_ja_{j})}\Big)\psi_{i_k}\ln(1+i_ka_{k}).
\end{align}
Note that \eqref{eq_average_highsnr1} is a weighted sum of terms of the form $\ln(1+i_ka_{k})$ for each transmitter  $k\in\mathcal{S}$ and each eavesdropper $n\in[N]$; collecting terms for each $k$, we may write 
\begin{align}
\label{eq_average_highsnr3}
&\mathbb E[C_s^{\mathcal{S}}]
= \sum_{k\in\mathcal{S}} \sum_{n=1}^N \ln(1+na_{k})w_{kn}(\mathcal{S}),
\end{align}
where $w_{kn}(\mathcal{S})$ is defined as 
\begin{align}
\label{eq_weights_S}
w_{kn}(\mathcal{S})&=\psi_n\sum\limits_{\mathcal{Q} \subseteq \mathcal{S}\backslash \{ k \} } \sum_{\substack{\mathbf{i}\in[N]^{|\mathcal{Q}|}\\\mathbf{i}=(i_j)_{j\in{\mathcal{Q}}}}}\prod\limits_{\substack{j\in\mathcal Q }} \frac{\psi_{i_j}i_ja_{j}}{(na_{k}-i_ja_{j})}.
\end{align}
}

 {
Finally, the ESR (\ref{eq_avg_capacity}) of optimal selection including backhaul uncertainty can be expressed in a simplified form as
\begin{align}
\label{eq_final_avg_sec_high1}
\bar{C}_s&= \sum\limits_{\mathcal{S} \subseteq [K]}\mathbb{P}[ \mathcal{S}]\sum_{k\in\mathcal{S}} \sum_{n=1}^N \ln(1+na_{k})w_{kn}(\mathcal{S})\\ 
\label{eq_final_avg_sec_high2}
&=\sum\limits_{k\in [K]}\sum_{n=1}^N \mu_{kn} \ln\lb(1+na_{k}\rb),
  \end{align}
where 
$\mu_{kn}=\sum_{\substack{ \mathcal{S} \subseteq [K], k\in \mathcal{S}} }  \mathbb P[\mathcal S]w_{kn}(\mathcal{S})$. 
}

 { We can observe from \eqref{eq_final_avg_sec_high2} that at high SNR, the ESR is the weighted sum of the logarithms of the (scaled) ratios between individual source-to-destination and source-to-eavesdropper SNRs. Also, the summation is over all transmitter-eavesdropper pairs. Due to the occurrence of the \emph{ratio} of $a_k$ values in \eqref{eq_weights_S}, we also conclude that these weights do not depend on the average SNRs, but only on the relative SNR strength between channels. Also note that the simplified expression in \eqref{eq_final_avg_sec_high2} captures the contribution of each transmitter-eavesdropper pair to the total ESR.} 

 { 
It should be noted that with suitable modification, the ESR expression given by   (\ref{eq_averaging2}) or (\ref{eq_average_highsnr3}) captures the special cases of optimal TAS  \cite{Sadeque_Subramanian_Average_secrecy_rate}, and optimal RS with DF relays \cite{Kundu_selection_GC16,Kundu_relsel_TWC15}. }
\vspace{-.5cm}
\section{Asymptotic Analysis}
\vspace{-.3cm}
\label{sec_asymp}
  {
In this section, an asymptotic analysis is presented
assuming the high-SNR approximation from the previous section. By \emph{asymptotic analysis} we mean the following: for each $k$, the eavesdropper channel's average SNR $1/\alpha_k$ is fixed, while the destination channel's average SNR is equal to a fraction $\rho_k > 0$ of the SNR $1/\beta$, i.e., $1/\beta_k = \rho_k/\beta$, and we consider the case $1/\beta\rightarrow\infty$. 
The asymptotic expression will be presented in a form following \cite{Wang_Yuan_Physical_Layer_Security}, although the system model here is completely different. Since the approximation $1+na_k\approx na_k$ becomes tight in the asymptotic scenario, the asymptotic expression for the ESR in \eqref{eq_final_avg_sec_high2} can be expressed by transforming it into a linear function of $\ln{(1/\beta})$ via 
\begin{align}
& \bar{C}_s
\approx\sum_{k=1}^K\sum_{n=1}^N\mu_{kn} \ln\lb(na_{k}\rb)=\sum_{k=1}^K\sum_{n=1}^N\mu_{kn} \ln\Big(\frac{n\alpha_{k}\rho_k}{\beta}\Big)\nn\\
\label{eq_esr_asy}
 &=S_\infty(\ln(1/\beta)-\mathcal{L}_\infty),
  \end{align}
where the high-SNR slope and power offset parameters are given by
  \begin{align}
\label{eq_inid_slope}
S_\infty&=\sum_{k=1}^K\sum_{n=
1}^N\mu_{kn},\\
\label{eq_inid_offset}  \mathcal{L}_\infty&=\frac{1}{S_\infty}\sum_{k=1}^K\sum_{n=1}^N\mu_{kn}\left(\ln\lb(\frac{1}{n\alpha_{k}}\rb)-\ln\lb(\rho_k\rb)\right), 
  \end{align}
respectively. 
In the high-SNR scenario, the slope and offset are the two important  parameters which determine the ESR. Note that for any $\mathcal{S} \subseteq [K]$, the sum of the weights $w_{kn}(\mathcal{S})$ over all $k$ and $n$ can be shown to equal unity, i.e., 
$\sum_{k\in \mathcal{S}}\sum_{n=1}^Nw_{kn}(\mathcal{S})=1$.
It follows that the high-SNR slope with unreliable backhaul is always less than unity, i.e.,
  \begin{align}
  \label{eq_slope_inid_high}
&S_\infty=\sum_{k=1}^K\sum_{n=1}^N\mu_{kn}=  \sum_{k=1}^K\sum_{n=1}^N\sum_{\substack{ \mathcal{S} \subseteq [K], k\in \mathcal{S}} }  \mathbb P[\mathcal S]w_{kn}(\mathcal{S})\nn\\
    &=\sum_{\substack{\mathcal{S} \subseteq [K] \\ \mathcal{S}\ne\emptyset}} \mathbb P[\mathcal S]\Big(\sum_{k\in \mathcal{S}}\sum_{n=1}^Nw_{kn}(\mathcal{S})\Big)
    =1-\mathbb{P}[\mathcal S=\emptyset].
\end{align}
 }
 {
 It can be observed from (\ref{eq_slope_inid_high}) that when the backhaul is perfectly reliable (equivalent to TAS), the high-SNR slope is unity. Furthermore, in this case the slope does not depend on the number of sources or eavesdroppers (this is also observed in \cite{Wang_Yuan_Physical_Layer_Security}), in contrast to the case when backhaul is unreliable. In the latter case, the slope depends on the probability of backhaul activity and is always less than unity.  It is noted that the high SNR slope actually represents the number of degrees of freedom or maximum
multiplexing gain \cite{Wang_Yuan_Physical_Layer_Security}. From \eqref{eq_inid_offset} we also see that the offset in unreliable backhaul is independent of $1/\beta$, however, it captures the effect of the destination and eavesdropper channels as well as the backhaul reliability, via $K$, $N$, and $\{\rho_k,1/\alpha_k,  \delta_{k} \}$ for $k \in [K]$.}
\vspace{-.5cm}
\section{Special Case: IID channel and Single Eavesdropper}
 {
In this section we consider a special case with a single eavesdropper, where the destination and eavesdropper channels are IID, i.e., $\alpha_k=\alpha$ and $\beta_k=\beta$ for all $k$ (however, we do not necessarily have $\alpha=\beta$). Here we will consider the high-SNR approximation scenario only. Since the ESR for this special case cannot be derived simply by equating all channel parameters in the solution of Section \ref{sec_opt_high} (due to the constraint that all $i_ka_k$ must be different in \eqref{eq_partial_frac_high}) with $N=1$, we derive the ESR for this case separately.} This case will yield a simpler analytical result which will bring more insight into the system behavior. Following a similar approach to that of Section \ref{sec_opt_high}, the CDF in (\ref{eq_cdf_inid_high1}) 
becomes
\begin{align}
F_{\Gamma_{\mathcal{S}}}(x) 
=\Big(1-\frac{a}{x+a}\Big)^{|\mathcal{S}|}
=1+\sum_{k=1}^{|\mathcal{S}|} (-1)^{k} \binom{|\mathcal{S}|}{k} \frac{a^k}{\lb(x+a\rb)^{k}},
\end{align}
where  $a={\alpha}/{\beta}$. 
Now $\mathbb{E}[C_s^{\mathcal{S}}]$ is evaluated using integration by parts as
\begin{align}
&\mathbb E[C_s^{\mathcal{S}}]=\int_1^\infty\ln(x)f_{\Gamma_{\mathcal{S}}}(x)dx
=\sum_{k=1}^{|\mathcal{S}|}(-1)^k \binom{{|\mathcal{S}|}}{k}a^k\nn\\
&\times\Big(\Big[\ln(x)\frac{1}{\lb(x+a\rb)^{k}}\Big]_1^\infty-\int_1^\infty \frac{1}{x\lb(x+a\rb)^{k}}dx\Big).
\end{align}
By taking the limit, the first term inside the brackets can be easily shown to be equal to zero. Moreover, using the partial fractions method for the second term and with some algebraic manipulations, we obtain
\begin{align}
\label{eq_iid_high}
&\mathbb E[C_s^{\mathcal{S}}]
=\sum_{k=1}^{|\mathcal{S}|}(-1)^{k+1}  \binom{|\mathcal{S}|}{k}a^k\int_1^\infty\Big(\frac{1}{a^kx}-\frac{1}{a^{k}(x+a)}\nn\\
&- \sum_{j=2}^{k}\frac{1}{a^{k-(j-1)}(x+a)^{j}}\Big)dx
=\ln(1+a)\nn\\&+\sum_{k=1}^{|\mathcal{S}|}(-1)^{k}  \binom{|\mathcal{S}|}{k} \sum_{j=2}^{k}\frac{1}{(j-1)(1+\frac{1}{a})^{j-1}}.
\end{align}
Finally, the ESR including backhaul uncertainty can be expressed with the help of \eqref{eq_prob_sel} and \eqref{eq_avg_capacity} as
\begin{align}
\label{eq_iid_high_back}
&\bar{C}_s=(1-\mathbb{P}[ |\mathcal{S}|=0])\ln\lb(1+a\rb)
\nn\\
&
+\sum\limits_{\substack{\mathcal{S} \subseteq [K] \\ \mathcal{S}\ne\emptyset}} \mathbb{P}[ \mathcal{S}]\Big(\sum_{k=1}^{|\mathcal{S}|}(-1)^{k}  \binom{|\mathcal{S}|}{k} \sum_{j=1}^{k-1}\frac{1}{j (1+\frac{1}{a})^{j}}\Big).
\end{align}

Rearranging (\ref{eq_iid_high_back}) assuming  \emph{uniform backhaul reliability} (i.e., $\delta_k = \delta$ for all $k$) in the form of \eqref{eq_esr_asy}, 
we obtain
the asymptotic high-SNR slope and power offset parameters as  
\begin{align}
\label{eq_iid_slope}
S_\infty&=(1-\mathbb{P}[ |\mathcal{S}|=0])=(1-(1-\delta)^K),\\
\label{eq_iid_offset}
\mathcal{L}_\infty&=
\ln\lb(1/\alpha\rb)-\frac{1}{S_\infty}
\sum_{k=1}^{K} \binom{K}{k} \delta^k(1-\delta)^{K-k}
\nn\\
&\times
\Big(\sum_{r=1}^{k}(-1)^{r}  \binom{k}{r} \sum_{j=1}^{r-1}\frac{1}{j}\Big)\nn\\
&=\ln\lb(1/\alpha\rb)-\frac{1}{S_\infty}
\sum_{k=1}^{K} \binom{K}{k} \delta^k (1-\delta)^{K-k}H_{k-1},
\end{align}
respectively, where $H_{k-1}$ denotes the $(k-1)$-th harmonic number.

\textcolor{black}{ The analysis with the IID channel condition and uniform backhaul reliability in \eqref{eq_iid_slope} reveals that when $\delta$ is small and $K$ is large, $S_\infty$ is approximately equal to $\delta K$, i.e., the expected number of active sources. Accordingly, the slope of the ESR curve will scale approximately in proportion to the number of sources and the backhaul reliability.}

\textcolor{black}{
 From the expression \eqref{eq_iid_offset} for the case of the IID channel condition, it is easy to show that an improvement in the eavesdropper channel quality increases the offset and hence degrades the secrecy performance. Similarly, an increase in $K$ and $\delta$ reduces the offset, and as a result the secrecy performance improves.}

\vspace{-.6cm}
\section{\textcolor{black}{Results}}
\label{sec_results}

\begin{figure}
  \centering
\includegraphics[width=2.8in]{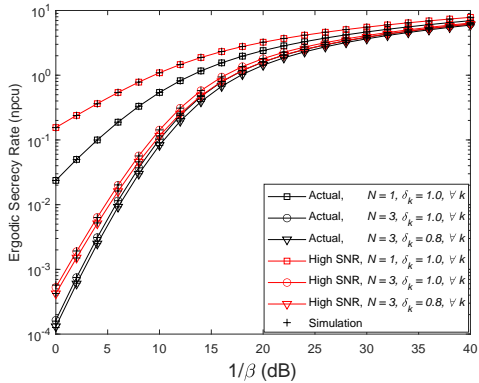} 
 \vspace{-.2cm}
\caption{ESR and its high-SNR approximation for optimal selection.}
  \label{fig_EXACT_HIGH}
  \vspace{-0.6cm}
  \end{figure}

 {In Fig. \ref{fig_EXACT_HIGH}, the ESR performance of the optimal source selection scheme and its high-SNR approximation are presented as a function of $1/\beta$ for different channel and backhaul reliability conditions.
We consider $K=4$, $N\in\{1, 3\}$, $(1/\alpha_k, \rho_k) = $ $(3\mathrm{ dB},0.1)$, $(6\mathrm{ dB}, 0.2)$, $(9\mathrm{ dB}, 0.3)$, $(12\mathrm{ dB},0.4)$, where $1/\beta_k=\rho_k/\beta$ for all $k$ \cite{Kundu_relsel_TWC15,Kundu_TVT19}.
We consider two cases: i) where all backhaul links are active (equivalent to TAS), i.e., $\delta_k=1$ for all $k$ and ii) when backhaul reliability is imperfect, i.e., $\delta_k=0.8$ for all $k$. }
As expected, the high-SNR approximation is actually an upper bound which approaches the actual ESR as the SNR increases. As the backhaul reliability improves, the ESR performance also improves.  { Furthermore, an increase in the number of eavesdroppers has an adverse effect on the secrecy.  }

\begin{figure}
  \centering
\includegraphics[width=2.8in]{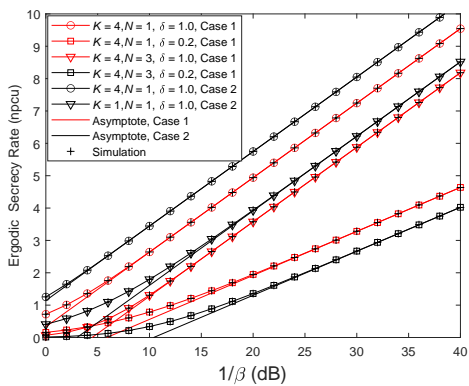} 
 \vspace{-.2cm}
  \caption{Asymptotic analysis of the ESR and its high-SNR approximation.}
  \label{fig_HIGH_ASYMP}
  \vspace{-.6cm}
  \end{figure}

Fig. \ref{fig_HIGH_ASYMP} presents the ESR along with its asymptotic expression for two representative cases (here the high-SNR approximation to the ESR is used). The ESR axis is plotted on a linear scale to show the asymptotic results as straight lines.  {In Case 1, we set $K=4$, $N\in\{1, 3\}$, and $(1/\alpha_k, \rho_k) = $ $(3\mathrm{ dB}, 1)$, $(6\mathrm{ dB}, 1)$, $(9\mathrm{ dB}, 1)$, $(12\mathrm{ dB}, 1)$  \cite{Kundu_relsel_TWC15,Kundu_TVT19}. In Case 2 (which corresponds to the special case in Section VI), we set $K \in \{1, 4\}$, $N=1$, $(1/\alpha_k, \rho)=(3\mathrm{ dB},1)$ for all $k$. In both cases, uniform backhaul reliability is considered with $\delta \in \{0.2, 1\}$.}
 {The performance can be seen to be improved by increasing $K$ for a given $\delta$;  improving $\delta$ for a given $K$;  degrading $1/\alpha_k$ from $\{3, 6, 9, 12\}$ dB for $K=4$ in Case 1 to $3$ dB in Case 2 for a given $\delta$; and decreasing $N$.}

 {If we consider the slope of the asymptotic curves, it can be seen that the curves with the same $\delta$ have the same slope irrespective of the values of $K$ and $N$. Further, the slope is significantly lower when $\delta=0.2$ compared to the case when $\delta=1$. It confirms that the slope does not depend on $K$ and $N$, and that the ESR performance is significantly affected by the backhaul reliability. We observe that when $K=4$, $N=1$ and $\delta=1$, although the slope is the same, the ESR in Case 2 is larger than in Case 1. This is because of the low offset  due to the low eavesdropping SNR $1/\alpha=3$dB in Case 2 compared to Case 1.
}

\section{Conclusion}
\label{sec_conclusion}
The ESR of optimal source selection in a multi-source multi-eavesdropper system with wireless backhaul has been evaluated in closed-form. A simplified analysis at high SNR, along with an asymptotic analysis, shows the effect of the backhaul reliability and channel parameters on the ESR.
We demonstrate that the slope of the ESR asymptote increases as the backhaul reliability improves. Our solution provides a generalized analysis of the ESR for optimal TS, optimal TAS, and optimal RS problems with or without wireless backhaul.
\bibliographystyle{IEEEtran}
\bibliography{IEEEabrv, COG_BACKHAUL}

\begin{thebibliography}{10}
\providecommand{\url}[1]{#1}
\csname url@samestyle\endcsname
\providecommand{\newblock}{\relax}
\providecommand{\bibinfo}[2]{#2}
\providecommand{\BIBentrySTDinterwordspacing}{\spaceskip=0pt\relax}
\providecommand{\BIBentryALTinterwordstretchfactor}{4}
\providecommand{\BIBentryALTinterwordspacing}{\spaceskip=\fontdimen2\font plus
\BIBentryALTinterwordstretchfactor\fontdimen3\font minus
  \fontdimen4\font\relax}
\providecommand{\BIBforeignlanguage}[2]{{%
\expandafter\ifx\csname l@#1\endcsname\relax
\typeout{** WARNING: IEEEtran.bst: No hyphenation pattern has been}%
\typeout{** loaded for the language `#1'. Using the pattern for}%
\typeout{** the default language instead.}%
\else
\language=\csname l@#1\endcsname
\fi
#2}}
\providecommand{\BIBdecl}{\relax}
\BIBdecl

\bibitem{Kundu_relsel_TWC15}
C.~Kundu, S.~Ghose, and R.~Bose, ``{Secrecy Outage of Dual-hop Regenerative
  Multi-Relay System with Relay Selection},'' \emph{{IEEE} Trans. Wireless
  Commun.}, vol.~14, no.~8, pp. 4614--4625, Aug. 2015.

\bibitem{Kundu_selection_GC16}
C.~Kundu, T.~M.~N. Ngatched, and O.~A. Dobre, ``Relay selection to improve
  secrecy in cooperative threshold decode-and-forward relaying,'' in
  \emph{Proc. IEEE Global Communications Conference}, Washington DC, USA, Dec.
  2016, pp. 1--6.

\bibitem{Shao_Modify_and_Forward}
S.~{Chu}, ``Secrecy analysis of modify-and-forward relaying with relay
  selection,'' \emph{IEEE Trans. Veh. Technol.}, vol.~68, no.~2, pp.
  1796--1809, Feb. 2019.

\bibitem{Kundu_TVT19}
J.~{Zhang} \emph{et~al.}, ``Secrecy performance of small-cell networks with
  transmitter selection and unreliable backhaul under spectrum sharing
  environment,'' \emph{IEEE Trans. Veh. Technol.}, vol.~68, no.~11, pp.
  10\,895--10\,908, Nov. 2019.

\bibitem{kim_Poor_Secrecy_full-duplex}
H.~{Liu} \emph{et~al.}, ``Secrecy performance of finite-sized cooperative
  full-duplex relay systems with unreliable backhauls,'' \emph{IEEE Trans.
  Signal Process.}, vol.~65, no.~23, pp. 6185--6200, Dec. 2017.

\bibitem{kundu_LSTM_GC20}
S.~Tripathi \emph{et~al.}, ``Recurrent neural network assisted transmitter
  selection for secrecy in cognitive radio network,'' in \emph{Proc. IEEE
  Global Communications Conference}, Taipei, Taiwan, 7-11 Dec. 2020, pp. 1--6.

\bibitem{kim2015security}
L.~Wang \emph{et~al.}, ``Security enhancement of cooperative single carrier
  systems,'' \emph{IEEE Trans. Inf. Forensics Security}, vol.~10, no.~1, pp.
  90--103, Jan. 2015.

\bibitem{kim2016secrecy}
K.~J. Kim \emph{et~al.}, ``Secrecy performance of finite-sized cooperative
  single carrier systems with unreliable backhaul connections,'' \emph{IEEE
  Trans. Signal Process.}, vol.~64, no.~17, pp. 4403--4416, Sep. 2016.

\bibitem{kim_Poor_Secrecy_Finite-Sized}
H.~{Liu} \emph{et~al.}, ``Secrecy performance of finite-sized in-band selective
  relaying systems with unreliable backhaul and cooperative eavesdroppers,''
  \emph{IEEE J. Sel. Areas Commun.}, vol.~36, no.~7, pp. 1499--1516, Jul. 2018.

\bibitem{Kim_Poor_Secrecy_CDD}
K.~J. {Kim} \emph{et~al.}, ``Secrecy performance analysis of distributed
  asynchronous cyclic delay diversity-based cooperative single carrier
  systems,'' \emph{IEEE Trans. Commun.}, vol.~68, no.~5, pp. 2680--2694, May
  2020.

\bibitem{Sadeque_Subramanian_Average_secrecy_rate}
N.~{Sadeque}, I.~{Land}, and R.~{Subramanian}, ``Average secrecy rate under
  transmit antenna selection for the multiple-antenna wiretap channel,'' in
  \emph{Proc. IEEE International Symposium on Personal, Indoor, and Mobile
  Radio Communications}, London, UK, Sep. 2013, pp. 238--242.

\bibitem{book_ryzhik}
I.~S. Gradshteyn and I.~M. Ryzhik, \emph{Table of Integrals, Series and
  Products}, 6th~ed.\hskip 1em plus 0.5em minus 0.4em\relax San Diego, CA:
  Academic Press, 2000.

\bibitem{Wang_Yuan_Physical_Layer_Security}
L.~{Wang} \emph{et~al.}, ``Physical layer security of maximal ratio combining
  in two-wave with diffuse power fading channels,'' \emph{IEEE Trans. Inf.
  Forensics Security}, vol.~9, no.~2, pp. 247--258, Feb. 2014.

\end{thebibliography}
\end{document}